# Models & Metaphors
## Part I: From the Boundaries of Formalization to the Visualization of the Non-Formalized

**Rainer E. Zimmermann[1], Sabine Ley**
IAG Philosophische Grundlagenprobleme,
Fachbereich 1, Universität,
Nora-Platiel-Str. 1, D – 34127 Kassel
e-mail: pd00108@mail.lrz-muenchen.de

**Vladimir G. Budanov, Vjacheslav E. Voitsekhovitch**
Institute of Philosophy,
Department of Interdisciplinary Studies,
Russian Academy of Sciences,
14 Volkhonka, R – 119842 Moscow
e-mail: nsavicheva@mtu-net.ru

## Abstract

The explicit relationship among perception, communication, and design is being discussed in some detail, in order to relate it to characteristic details of the modelling of the world which defines the scientific and artistic activities of humans, in the first place, but also the philosophical grasping of the world and its foundations in order to position humankind within this worldly framework, at the same time. In particular, the structure and constitutional role of language are being analyzed with a view to the latter's universality. For the special case of synergetics (within the recent context of theories of self-organization and the formation of structure), this cognitive background is being inspected as to possible consequences for aspects of complexity and non-linearity.

## 1.  The Relationship Among Perception, Communication & Design

It has been shown at other occasions [2] that recent results of modern physics can be used to shed some more light onto the foundations of the world, provided the actual task of philosophy is being re-interpreted in terms of a theory which is following up the results of science rather than laying the grounds for the latter, contrary to what the original intention of Aristotelian *prima philosophia* would imply. As it turns out, the interpretation of the main results of present research dealing with aspects of quantum information theory and quantum gravity, respectively, as well as with self-organized criticality, suffices to re-construct a large class of phenomena not only within the field of physics proper, but also within chemistry, biology, and even the social sciences.

---

[1] Also: Clare Hall, UK-Cambridge CB3 9AL. Present Address: Lehrgebiet Philosophie, FB 13 AW, FH, Lothstr.34, D – 80335 Muenchen

As seen under a *philosophical* perspective, it can be shown that the general conceptualization of such a unified view of the world has been prepared on a long line of thought which begins with the Greek Stoá and leads up to the theories of Spinoza, Schelling and Bloch as some of its representatives, eventually showing up in a somewhat modified form in what can be called *transcendental materialism* today. [2] As seen under the *physical* perspective however, it can be demonstrated how a philosophy re-interpreted in the above-mentioned sense can unfold a heuristic potential which is able to produce guidelines of orientation as to deciding about competing concepts in the sciences. Such a function of philosophy has often been asked for, but has rarely been realized so far when there actually was one of the few occasions to explicitly deal with it in detail. One problem of this is to sort out what would be the appropriate philosophical approach to pertinent questions, in the first place. In the present paper we shall try to contribute somewhat to a further clarification.

But although being oriented with a view to the sciences, philosophy is not at all duplicating the latter's work. Contrary to that, philosophizing means primarily to critically reflect each single sector of the uncovered worldliness with a view to the global interrelationships among all possible sectors, to ask for necessary and sufficient conditions for the main structural aspects of single sectors, and for possible alternatives. In this sense, philosophical reflexion gains more and more the aforementioned *heuristic* connotation. The unification of the scientifically (necessary) segmented actuality is no self-purpose therefore: On the contrary, the important point is to show, within all what is empirically observable, what *is not more* subject of science.

We have also seen at an earlier occasion [1] that a clear distinction between the (physical) world and its foundation is one of the central aspects that have not very often been the topic of conceptualizing ongoing research in the sciences. The same is even more pertinent as to the *holistic* aspects of this approach: The point is that if we visualize some abstract mathematical structure as representing what we can infer about some suitable foundational basis of the world, provided consistency is being secured in terms of being able to actually derive what we know about the world from that structure, then this representation serves as a mode of description of all what there actually is in worldly terms, i.e. it is not only referring to the world as it can be described in terms of physics, chemistry, and biology (or computer science), but also in terms of psychology, sociology, economics, and everything else. In other words: It is our usual perception of everyday life mapped as a gross average of all what there is that is nothing but a collection of aspects of the (universally) underlying being of worldliness as it can be visualized under the perspective of the human mode of being. And within the course of communication, these aspects are permanently phrased into a propositional form which is governed by the appropriate language chosen. Hence, the ordering of available modelling procedures is itself a *linguistic* aspect. Such fairly traditional insight of philosophy [2] is however alien to modern science, despite its formal language (mathematics) being only a special case of the aforementioned. This is mainly so because the primary task of science is to deal with the world as it can be perceived (observed). Hence, science is exclusively aiming at the empirical world, not at its foundation. And this empirical world is visualized moreover as a stratified set of regions accessible to various, more or less disjoint, disciplines.

The important consequence is after all, that the modelling (i.e. observing, conceptualizing, and re-constructing) of the world itself turns out to be part of the process it is going to model in the first place. In other words: If we call the substratum of the world „matter" (or more precisely: „space-time-matter"), then *thinking itself* is nothing but a form of this matter. It is probably a very complex form as to that, but it is basically nothing else than that sort of matter which is available from the beginning on, and this is *physical* matter. However, as we have seen before, though the world can be visualized in terms of physical matter, matter is not already substance. Instead, substance is the world's (and hence matter's) foundation. Nevertheless, if thinking of the physical world, we are basically thinking of its „framework categories" which are space, time, and matter. And those are the categories with respect to which we actually phrase our propositions, in the first place.

Note that what we advocate here is not necessarily a sort of classical reductionism, nor is it something which could be summarized as „naturalism" or mere „materialism": First of all, we have to take care of the differentiating power of our arguments. (If everything is matter and/or nature, then nothing is.) Hence, the crucial point is that each level of complexity (of the observable world) has a common boundary with some higher level, as well as with some lower level, but that obviously, all these (worldly) levels of structure are *irreducible* with respect to each other. [3] This is so because lower levels are „sublated" within higher levels in the threefold Hegelian sense of the meaning of „to sublate" (namely: to raise, to conserve, and to abolish, at the same time). In this sense, as Thomas Kuhn has shown some time ago, Newtonian physics e.g. shows up as a special case of relativity, if a certain limit of the latter is being taken in terms of parameters which are part of the theory. But in *cognitive* terms, this is a different outcome as compared to the period when relativity theory was yet unknown. Both cases (visualizing Newtonian physics as a special case, and visualizing the whole of mechanics in terms of Newtonian physics) differ in their space of implications. The latter has been enlarged by now. Hence, their associated viewpoints are irreducible.

The reason for this is mainly that humans have to think sequentially – collecting one argument after the other. They are *digitalizing* the world which primarily shows up as an analogous process in order to be able to conceptualize it. Irreducibility therefore, is a cognitive analogue of temporality. This is also the reason for calling the materialism put forward here „transcendental materialism", because the word „transcendental" implies that the substratum discussed is not al-ready substance itself: The foundation of the world is not accessible to an inhabitant of that world, due to the fact that perception is restricted. Moreover: The observed process (of evolution) can be thought of as a „self-process" of the underlying foundation. Or rather: as a *self-explication* of the latter.

On the other hand, second, when all of this stuctural hierarchy is, as we know by now, simply a property of human perception according to which the world is actually being modeled, then in principle, the world as being visualized in terms of a complex hierarchy of forms is nothing but the (epistemological) representation of what humans can grasp of the underlying substance, given their perspective which is developed according to what they can perceive. Hence, instead of being an indication of the world's complexity, the observed hierarchy of forms is nothing but an indication for human reflexion's tendency to actually *simplify* what it perceives. Despite the obvious complexity of the world therefore, this world is, as it turns out, much less complex than its foundation. The point is that reflexion means *reduction of complexity*, in the first place. [4] And this can be visualized as the characteristic human activity indeed: to model the world in a simplified way in order to make it tractable for human *praxis*.

So, in a way, everything starts with language. Or to be more precise: It starts with the communication of (reflexive) thoughts and its mapping in terms of appropriate scripture. If we visualize a language as a collection of lexicology, syntax, and semantics, then we have to look for the correspondence between its phonology and its graphology first. We can think of oral language as a category ORL whose objects are phonemes and whose morphisms are rules that produce chains of phones (tones) which are meaningful. Correspondingly, scripture (written language) is a category SCRPT whose objects are graphemes and whose morphisms are rules that produce chains of letters. Writing then, is simply a functor ORL $\rightarrow$ SCRPT. There is also a formal correspondence between the detailed structuring of the morphisms in ORL and that in SCRPT being cared of by this functor. Additionally, there are some more criteria of form such as eugraphical and/or rhythmic aspects (of syllables e.g.), and there is a characteristic sub-structuring of the respective lexicology, syntax, and semantics. While the first (lexicology) is dealing with the form of words (conjugation) and with their functions (tempus, modus, genus), the second is simply a system which associates structure descriptions with propositions. In other words: A syntax is a system of rules which models the consistent distribution of word forms. Referring to earlier work by Chomsky [5], we call a grammar *generative*, if it provides for the consistent association of three basic components: syntactic (relating to information which is necessary to identify and/or interpret propositions), phonological (relating to information necessary to interpret the phonetic structure, thus mapping the syntactic structure to the phonetic), and semantic (mapping the syntactic structure to the appropriate meaning). Hence, it is the syntactic structure of a language that specifies its deep structure (as to its semantic interpretation) as well as its surface structure (as to its phonetic interpretation). Note that „interpretation" here is still referring to the technical aspects of propositional ordering, not yet to the full *hermeneutic* process, although the former is obviously being placed at the beginning of the latter. The reason for this is mainly that the ordering principles in the sense of the syntax have to be checked in the first place, before one would be able to actually extract the (probably) correct meaning of what has been said or written. In order to display the systematic of a language in more detail, there are well-known examples of graphic mappings which represent the detailed structure of a syntax as introduced by Chomsky in terms of suitable tree diagrams ([5], p.90)

Phrase markers of this Chomskyian type indicate that the syntax can be mapped in detail by means of directed graphs, i.e. in terms of a category PHRS whose objects are vertices and whose morphisms are edges. In other words: There will be a pair of functors ORL → PHRS, and SCRPT → PHRS such that both the underlying systematic of the respective categories as well as their semiotic contents are being represented in terms of appropriate digraphs. (Note that the symbol strings as those utilized for scriptures themselves are also contained in these functors.)

We cannot develop a whole theory of languages here by means of categories (this will be actually done elsewhere [6]). But the important point for us is to notice that while we are able to represent linguistic aspects in categorial terms, we can also utilize what we know about their underlying logical consequences. So in the following we concentrate on the interpretational procedure as it is related to the analysis of propositions. We use the aforementioned as an insight into the fact that modelling is essentially based on a systematic of predication. However, we will not deal with more foundational aspects of linguistics, as they are being advocated by Chomsky. In particular, we do not share Chomsky's assumptions on the universality of generative grammars. [7] (For a more detailed description of linguistic principles refer to the standard volume of Lyons. [8]) But what we do is to stay with the logical consequences of interpretations. This has been discussed in more detail elsewhere [9], [10], when referring to a suitable approach to formalizing the transition from logic to hermeneutic demonstrating at the same time that the latter can be visualized as a generalized version of the former, for the case of incomplete information.

In this sense, the description of processes can be mapped in terms of propositions formulated in some language. Hence, under this perspective, languages altogether can be thought of as being categories whose objects are propositions formed out of a given generative grammar including some suitable lexicology, syntax, and semantics, satisfying rules as laid down above, these rules being implicit in the category's morphisms. *Translation* then is a functor between language categories mapping objects to objects (propositions to propositions) and morphisms to morphisms (rules to rules). The idea of this approach is that translations are path-dependent with respect to compositions. This can be shown in that the generic diagram representing the form of mappings,

$$A \rightarrow B$$
$$\downarrow \quad \downarrow$$
$$C \Leftrightarrow C,$$

is not commutative, if we deal with any natural languages (A, B, and C, say). (The double arrow indicates the identity mapping.) But *it is* commutative, if we deal with logical languages (i.e. formal languages with different underlying logics). What we can say is that in the case of the logics, the information (with respect to the semantics of the propositions utilized, e.g. with a view to applications in physics) is (almost) *complete*. Hence, translation of propositions is commutative in terms of a suitable diagram representation (as shown above), if we deal with logic. It is thus *path-independent*. But for a hermeneutic approach, which shows up here as a generalized logic for incomplete information, it is *path-dependent*. (To turn this the other way round: Logic is a hermeneutic with complete information. Each hermeneutic has thus its logic nucleus: This means nothing is completely arbitrary. On the contrary, there is always a rational aspect which is at the foundation of what is actually being said about the world.)

The important insight gained here is that the world shows up as one which is inherently „questionable" in the sense that it is offering itself to hermeneutic interpretation from the beginning on. And it is this „beginning" which is initialized in terms of (human) perception. As has been shown earlier in some more detail [11], the basically anthropological foundation of the mode of perception turns out to be the underlying motivation for developing the viewpoint of what is referred to as „transcendental materialism". The idea is that perception is not primary in this sense, but it is „preformatted" according to the biological evolution which has been actually taken by humans. In other words: If perception is a product of the process which is being perceived afterwards, then the correspondence between that perception and what is being perceived is very close in the first place. Probably, one can speak of a kind of „fine tuning" which means that not very many variations of this „preformatting" appear to be possible. So what we assume is that not only is perception tuned to the perceivable world, but also is the mode of modelling the world as a „cognitive work" operating on what has been perceived. That is, all means of modelling, such as the pertinent graphism admitting of diagrammatic representations of models, are actually included in this cognitive fine tuning of a product of the world to the rest of this world.

While we postpone the explicit graphism for a while, we concentrate first on the general semiological aspects of perception: As has been shown by Leroi-Gourhan [12], it is the co-evolution of language and tools technology which results in an epistemological unification of gesture and tool by means of producing an explicit *sign structure* of the world. Signs turn out to be the basic elements of reflexion, because they combine the task of reflexion, i.e. the abstracting of symbols from concrete reality in order to constitute a parallel world of language which is able to act upon reality more effectively, with its communicative aspect showing up as its narrative structure. The framework categories for this unifica-tion are space and time: In fact, the association of temporal rhythms with spatial structures can be interpreted as a domestication of space and time. Social space therefore, is gaining the connotation of a mapping of space-time geometry in metaphorical form. (Most critics of this philosophical approach, particularly those from the fields of science, forget to take this „metaphorization" into account when discussing formal analogues as they are utilized sometimes in the philosophical field.)

In so far the *cognitive* aspect of human beings is showing up as an immediate expression of this aforementioned co-evolution. Piaget has already pointed to the utilization of category theory [13] when showing that perception and cognition are based on what he calls „pre-categories" derived from „rough systems of morphisms" in a permanent struggle between invariant forms and transformations. This is the reason for the structural morphisms which establish a hierarchy of mediated functors and functor categories to also carry their own internal logic. In other words: The process of signification itself shows up within an implicit double connotation of representing the intendend structure on the one hand, and of contextualizing it within a categorial framework of logical consistency on the other hand.

This has been the basic motivation for discussing consciousness itself in terms of geometric and/or computational metaphors. In a sense, we can visualize the process of modelling the world according to what can be perceived as a kind of „global algorithm" which is implemented into the world from the outset. On the other hand, this global sign structure of the world which is being re-processed all the time within the framework of the fundamental categories of space and time (note the double meaning of „categories" here which parallels the corresponding double meaning of the word „topoi") can be thought of as being mediated „down" into the local macro- and into the individual micro-environment in terms of „local algorithms" with respect to chosen strategies of modelling. Hence, „metaphorization" of the world is the underlying process of its understanding according to the explicit utilization of local strategies of hermeneutic signification. (In fact, as it turns out after all, „hermeneutic signification" shows up here as a tautology rather.)

The fact that the world presents itself as something which is „questionable", as we have mentioned before, shows up in the micro-detail of the process of signification which is underlying the communicating of the results of modelling the world, in terms of a „gap" between the signifier and its associated significate. This is what Lacan shows when deriving the process of signification from its linguistic elements. [14] The signifier is with respect to the significate in a relation of type S/s, where the „slash" refers to the fact that there is always a non-deletable gap between the one and the other such that the signifier is never capable of exhausting the exact meaning of what is being signified. (Note that the significate is thought here in terms of phonemes rather than of graphemes.) A metaphor then, is a kind of transformation of signifiers, written in Lacan's terminology as S'/S x S/s. Hence, „understanding" what is being said means ba-sically „transforming metaphores" within a given context. As Lacan discusses in more detail, this procedure is at the foundation of most of the psycho-analytic phenomena. We will not go into these details here, but the important point for us is that language is thus centred around interpretation, and that social phenomena (which are always phenomena of communication) derive from that „questionability" of the world which can be operated upon by modelling it, but which can never be solved completely. Hence, speaking about the world (as well as symbolizing/signifying the structure of the world), visualized as an interpretation which because of its communicability is always *translation* at the same time, shows up as permanent mis-representation/distortion or transposition (in Freudian terms: „Ent-Stellung"). *And this is what causes social phenomena.*

But there are two further consequences of this: First of all, this process of permanent mis-representation (in the procedure of representing) is the social basis for innovations. This is so, because each new metaphore acts as an „impertinent predication" onto the body of traditional language, and thus unfolds its heuristic potential. ([15], pp. vi, 87.) In this sense, the metaphore is to common language (including poetry) what the model is to science. ([15], p.228) It is no coincidence that innovative research can be related in this way to innovative thought in general. [16]

A second point to that is that the same process is also at the foundation of the permanent differentiating of differences which is determining the perception within the world. Deleuze has noted that to seperate substance from attribute (in the Spinozist tradition of „omnis determinatio est negatio") is possible by abstraction only. Hence, what is being expressed is also veiled at the same time and needs therefore, further interpretation. [17] In this sense, „expressio" comprises always of both „explicatio" and „complicatio". This is the reason for the fact that traditional metaphores of expression (such as „mirrors" that reflect or „germs" which are expressed in unfolding) point from the beginning on to graphic means of communication. More recently, Louis Kauffman has presented similar (though mathematically more rigorous) arguments in favour of founding the world (spacetime) on constitutive differentiations. [18] In the theory of differentiation as it has been presented by Derrida [19], the same aspect is showing up, because in the double meaning of the word „différance" (from: *différencier* and *différer*, at the same time) the process of producing differences and the process of delaying (suspending) the presence (of structure) which is actually being produced by this differentiating, is thought of as being set into one. In this sense, the active articulation of space-time is included in the structure of the discourse as it can be described in linguistic terms.

This can be hardly made clearer than by utilizing the insight gained from the anthropologically underlying *graphism* which is intrinsic to all forms of discourse. As Leroi-Gourhan has shown, the emergence of graphic symbols at the end of the *paleoanthropus* period can be interpreted as an indication for a new relationship between the cooperative poles of hand and face. ([12], ch.VI) The important point is that the graphism is emerging however from abstraction rather than from naive actuality. In other words: Graphism does not mean primarily the mapping of concrete forms. Instead, the realism of mappings is a late product of a long development. Graphism means also an unfolding of three new (spatial) dimensions as compared to the one (temporal) dimension which is open to phonetic expression. Much later then, it is the *linearization* of the graphism which actually creates scripture, but which restricts again this recently gained expressive freedom. Hence, the basic idea is to visualize the first graphism as a mode of expression which opens up a true parallel (of equal weight) to phonetic language. In this sense, the graphic result is *mythographic* rather than significative. The gesture interprets the (spoken) word, and the word comments on the graphism. But it is the (linear) graphic mapping then which leads to a consequent regularization and thus restriction of symbols. Grammar's time is approaching. And here is the root of rationality, because restriction means to actually restrict the expressive means for the irrational moments of life, in the first place.

This is the reason for the materiality of signification: The application of the hands in actually producing something can be expressed in terms of gestures which characterize the „language of the hands". (This function of the gesture has not been lost until today.) Its analogue is the (spoken) „language of the face". And the first graphism tries to express the gesture in terms of a spatially fixed *copy-picture* (or mythogram). Modern scripture and graphic mappings however, aim at the transition from the mythographic gesture to the formalized gesture. Standardization of expression, though connected with explicit restriction, opens up new horizons of communication. Hence, the universality of diagrammatic applications, not only within the framework of formalized languages: Graphic design is always representation. But it is, though strictly regularized very often, a formal translation of the material gesture. The gap between object and mapping cannot be overcome, but abstraction secures the universal applicability of this modernized (linear) graphism. And, as we have seen, this abstraction is always of equal weight than the original concretization.

Those are the basic ingredients of assembling knowledge by means of modelling the world: It is the intrinsic graphism which is securing the relevance of the outcome. We have a recursive systematic of producing structures by means of modelling and designing (i.e. re-constructing) them in graphical terms: By doing so we do not leave the cycle of abstraction. We rather circulate them under the formal and the „natural" (i.e. hermeneutic) perspectives.

What is being transported via the route indicated here is the aforementioned „expressio" which shows up as both „explicatio" and „complicatio". The empirically found is modelled in terms of physics utilizing the appropriate mathematical language. The latter is primarily based on a formal logic which can be expressed in terms of an adequate graphic design. (One has only to think of recent diagrammatic techniques as introduced by knot theory.) Intuitively, this skeleton of a graphism (heavily restricted by precise rules) can be visualized as a rudimentary kind of geometrization: Essentially, the utilized graphs do not only depict spatial and temporal dimensions, but they also transform one type of dimen-sion into the other. Categories (and more recently, topoi) are examples for a static mapping of dynamical phenomena. Philosophy then, is the field which transforms a formal input into a hermeneutic output in doing essentially two things: first, isolating basic (epistemic and thus cognitive) structures of scientific fields in order to unify them into one conceptual whole which can eventually serve as a presentation of the present totality of the world (this is what science cannot do), second, delimiting the knowledge gained against what is beyond science in principle, thus opening up the perspective as to what is discovered once the restriction of the formal discourse are lifted somewhat. The latter is what we call „foundational activity" of philosophy. The resulting generalized models are rephrased then in terms of a path-dependent, hermeneutic logic of translations (of meanings). Mathematics is being replaced then by semiology, geometrization is replaced by metaphorization. In the fashion of a somwhat „clockwise motion" we can indicate an increasing context-(and thus path)-dependence of argumentation. The important point is that although this motion of arguments looses more and more its formal strictness by means of the increasing relevance of the metaphorical design, it nevertheless is capable of an innovative, heuristic injection of argument once the results are being fed back into the available theories. This is the reason for always having a mixture of these different types of argument in physical theories rather than pure mathematical arguments only. (A detail Feynman has actually noticed a long time ago. [20]) This recursive process of producing knowledge has been discussed in more detail elsewhere. ([9], first reference) But for us here, it suffices to notice the relevance of the graphism intrinsic to the process of producing knowledge.

Starting with what we have learned from dealing with the generic difference between world and foundation, and thinking of the fact that it is human reflexion and representation which is actually modelling this difference in the first place, we can straightforwardly conclude that consciousness, as it is being conceptualized according to the same sort of modelling, shows up as a classical and thus emergent property of the world. This rules out any direct quantum approach to consciousness, because the concept itself is primarily defined in terms of the „experience" humans do actually have of it. The quality of experience however, is a conscious correlate to the brain; and the brain is a macroscopic object. This does not mean that the brain's macroscopic behaviour would not be determined by its microscopic and thus quantum level. But on that level it cannot be experienced, because human perception (as we have seen) is restricted to the classical level. Also, it cannot be experienced on the fundamental level, because in being the foundation of the world, it is not its proper part. It is rather to the world what possibility is to actuality. Nevertheless, as a possibility, consciousness is inherent of course to the foundation. But this does not mean that the fundamental level is „thinking" as we know it. Hence, any kind of universal pan-psychism has to be ruled out as to that. (Note by the way that human reflexion is the necessary but not sufficient condition for performing actions. The initiation of actions being undertaken by humans has always a macroscopic quality to it. That is, ontologically, humans are (classical) parts of their (classical) macro-world. On this level, their models of the world, although remaining models all the time, become concrete.

From the beginning on, it has been the main problem of talking about consciousness to define an appropriate relationship between consciousness and brain. The easiest approach was to visualize this relationship as an analogue to the relationship between software and hardware. Hence, the *computational metaphor* was being created. By many authors, its difficulties have been located in the problem of symbolic representation. [21] As we have seen earlier, within the onto-epistemic view, perception and cognition are closely tied to a generic graphism which is in turn determining the non-separable relationship between reflexion and communication (as well as action). On the one hand, this can be shown when talking about cognitive categories in terms of their being represented by attractors of connected (and massively parallelly organized) dynamical systems. ([21], p.5sq.) Because the model of attractors is itself a representation which can be translated into a suitable graphism. Hence, symbols are mediated all the time in terms of other symbols, even if on a very basic (if not fundamental) level, symbols and logical operations are difficult to locate. ([21], p.8)

Visualizing consciousness as an emergent macro-structure of the classical world, its „software" is to that of the fundamental level what the user language is to the machine language. The relationship between the two would be compatible with the recursive modularity of the brain as it is being discussed in terms of a self-organizing process according to the principles of self-organized criticality (SOC), comparable with Conway's „game of life" utilized for a model of the edge-of-chaos type. [22] The proposal is here that this type of dynamics is self-similar across multiple scales of neural organization (as it is usually the case for systems in SOC), utilizing a limit-cycle attractor for recognition, participation, and engagement states, while utilizing a chaotic attractor for ready, receptive, and disengaged states. ([22], p.57)

There is however, a complicated problem of self-reference when dealing with necessary conditions for consciousness: What we do just now is to think of humans as a product of a natural evolution which has been actualized all the necessary conditions for the possibility of eventually emergent human beings, starting from what we call the „Big Bang" until „recently" (i.e. some 100.000 years ago). In other words: There must be galaxies and stars, planetary systems as well as appropriate physical conditions, first, before it is possible to eventually actualize human beings on various planets. Hence, the condition for having consciousness as we know it, is the *complete history* of the Universe itself. (In fact, this viewpoint is the basis for deriving a principle of cosmological selection then, as Smolin has done.)

But on the other hand, *that this is really the case*, is an assumption of telling this story in the first place, i.e. of modelling the world such that this outcome is being achieved at all. And we know already that we do our modelling according to what we perceive and cognitively grasp. We have to admit therefore, that the picture developed cannot be correct in absolute terms. Because, space and time have been shown to be framework categories of human thinking. But if so, then in modelling the world, it is simply *pretended* that the world has developed in the aforementioned way, because this assumption is fitting well to what we can observe. But what we can observe, is not the whole story! We do *as if* the world develops within evolution, governed by the categories of space and time which enter explicitly the formal representation of our models. Doing so works well as far as a wide range of applications is concerned. But as to a discussion of the underlying foundation of the world, this is not really telling much.

We arrive now at two (short) conclusions of this. The first deals with the philosophical consequences as to human orientation within the world. The second deals with the ethical implications of this.

First of all, it should be noticed that we have not actually *shown* the derivational dependence of all what there is from its underlying foundation. What we have done instead is to *argue* in favour of the principles of this dependence and then to look for indications that demonstrate the plausibility of their assumption. Hence, we have basically argued in a *heuristic* manner, and this is essentially what the philosophical approach to science, the taking in sight of what physics has to offer, actually means.

Note that we have not argued in the manner of what is usually referred to as „analytical philosophy", because despite our interest in language, we have not forgotten that there actually *is* something about which language speaks. The point is that philosophy of science cannot be based on the available (scientific) theories alone, in so far as they are simply sets (or categories rather) of propositions which satisfy certain pre-defined rules. More than that there is also a „metaphysical" base point from where we have to argue, thus projecting an intrinsic (if not generic) systematic framework onto the aforementioned theories. This base point in turn, is derived from general rules of philosophical thinking, in the first place. The „theorem of sufficient reason", „Ockam's razor", or simply, the „principle of thought line consistence" ([9], first two references) are only a few examples for these rules. Another one is the primary difference between world and foundation. Hence, although these general rules do not actually interfere with ongoing practical research, they nevertheless re-structure the actual set of (still heuristic) starting points for any scientific enterprise. (This is, as mentioned before, what Feynman had noticed some time ago.)

What we have actually *shown* however, is that these heuristic aspects emerge from a line of thought which characterizes aspects of substance metaphysics. Obviously, this must be a modern sort of metaphysics today, as we cannot go back anymore to the state of knowledge as it was common some 300 years in our past. Hence, and we have argued thus, it is „transcendental materialism" which points to an innovative approach toward these aspects. Nevertheless, one basic result of ancient metaphysics has not changed at all until today: This is the fact that once we accept its underlying assumptions, we have to admit their universality for all what there actually is in onto-epistemic terms. Of course, kno-wing that physics is at the „bottom" of all what there is including everyday life in all its detailed aspects, does not save us having to do the „hard work". That is, it will be extremely unlikely that one day we might develop some sort of „behavioural physics" (as Comte originally thought 200 years ago). We will have to describe all the various fields according to their appropriate level of complexity in their own language, and we cannot expect to eventually replace psychoanalysis or economics by physics or quantum gravity.

But the insight we actually gain by knowing this, is more a heuristic insight into what kind of orientation humans should be involved in the future. Hence, this insight develops an eventually normative function, and leads therefore to a reconsideration of ethical aspects of human life. According to an expression of Hogrebe, we can call this a „fundamental heuristics." [23]

But note also something else: In a completely different field, namely that of urban planning, it has been shown recently how the holistic conception of practically „embedding" the intended topic into its structural background based on elementary principles of evolution deriving from a suitably chosen foundation is not only present in every technical detail of daily life, but also unfolds normative power within its framework. [24] In fact, as it turns out, the results exhibited by a close analysis of the various aspects of city life point towards what we may justifiably call „metaphysical implications". This viewpoint is still not very common within European science. On the other hand, nobody (including most of the scientists) would doubt the justification of Asian „metaphysical" aspects of architecture when applying principles such as those of the *Feng Shui* to the actual erection of buildings. (In fact, the Needham institute in Cambridge (UK) is built according to these principles.) And it is not very difficult to recognize harmonizing aspects within Asian city structure such as displaying the existence of „difference within a framework of repetition". However, as Henri Lefebvre has pointed out some time ago ( cf. [25], pp. 157sq.), this Asian structure of social space has one serious drawback: it is an *attribute of power*. It implies and is implied by divinity and empire, and it thus combines knowledge with power. Nature (whether available or constructed) is harmonized according to a hierarchy of principles which explicitly centres around the emperor, i.e. the representation of a feudal system. Obviously, this is something which explicitly contradicts the insight gained in the period of European enlightenment. Hence, contrary to the Asian approach, introducing the *concept of de-centralization* into the city structure reflects a fundamentally European approach so as to deal with these „metaphysical" aspects of daily life. Indeed, as we can find out when analyzing various fields in detail, de-centralization is not only a secularized principle of social organization, but also a basic principle of evolutionary processes abundant in nature (if visualized in terms of modern European science). This illustrates clearly that ethical implications of this „metaphysical" background are always present: In fact, we realize this in the very catalogue of (technical) measures taken when thinking of the underlying objectives showing up within the details of urban planning. It is not a coincidence that these implications are tied to legal aspects of democratic policies such as defining technical rights (e.g. of pedestrians) with respect to the human rights convention.

The point is that the decision for utilizing scientific methods is from the beginning on a decision for choosing an *onto-epistemic* approach to the world. Hence, it is also a decision in favour of acknowledging a rational nucleus of all what humans can perform within their environment. This aspect is important when thinking of the fact that contrary to what is usually claimed, political decisions are almost always based on emotional and explicitly irrational rather than on rational aspects. Nevertheless, ethics is the „science" (if you like) of what is adequate. This is the difference between ethics and morality: The former tells us what is adequate or not according to what we presently know, the latter values what is good and bad according to what we presently believe. Hence, politics, or rather its practical application in daily life in terms of elections, planning, executive actions and so forth (we are only talking about political systems which are based on principles of European democratic, legal, and republican thought), is usually judged following the prejudiced structure of traditional morality, and not, as it should be, the objective structure of rational ethics. The point is that the ancient problem of the Big Revolutions of the 18th century (the American, and the French), namely the transformation of the *bourgois* into the *citoyen* has not been solved lately. But if this cannot be achieved, it could be possible perhaps to approach a more modest objective: namely to transform the respective institutions instead, especially, if they are already of the democratic, legal, and republican type.

But beyond this layer of everyday politics, there is another, deeper layer of the intrinsic mediation of the ontological and epistemological levels of modelling the world and pratically behaving accordingly. In fact, if the Santa Fe school, or some of its individual protagonists, or protagonists from its vicinity, like Stuart Kauffman, Per Bak, and also Lee Smolin, talk about evolutionary principles on a fundamental level of nature (of which humans are a product), then what they actually do (perhaps without realizing this very clearly) is to aim at this intrinsic mediation and to connect it to an explictly ethical perspective (though the outcome might be different as can be seen by the detailed views of all of them). Hence, to visualize, as we put forward here, the world and everything what there is (including the city as a special example) as an emergent computational system consisting of massively parallel microworlds eventually creating an observable macroworld, means to apply the computational paradigm in an ethical perspective: If there *is* a computational paradigm, then there is a realistic possibility for an algorithmic approach to the world. It is mathematical logic in fact which provides this algorithm. And although we have to accept that many aspects of human life within social systems cannot be completely modelled according to mathematical procedures, what we can do nevertheless, is to notify the logical nucleus in all what there is, and this is the rational nucleus, at the same time. Hence, following a systematic pattern in scientific approaches means to take this rational nucleus explicitly into account. And this choice unfolds a whole bundle of ethical implications which are based on a holistic background of ideas. And because this background produces an ethical perspective, it has itself explicit normative qualities.

## 2.  The Universality of Language

The aforementioned has clearly illustrated the inherently universal character of language, not only as a defining quality of humans in anthropological terms, but also as a universal means of communicating the results of worldly modelling into the human community in order to eventually develop possible lines of applications (up to explicit strategies of complexity). Ley and Beller have dealt with a consequence of this in more detail, as to the possibility of mapping the universal character of language equally well by developing a universal language in the first place. [26] This ancient human project points to the following problem: Is there by any means, a globally invariant structure of language such as to represent a universally stably core of „langue" in the sense of Saussure? If so, this could be utilized as a starting point for developing a mean strategy of translating philosophical concepts into a truly *inter-cultural* context. The idea is then, to isolate a number of meanings which transcend all possible local differences of language, in the sense of valueing them as an „anthropologically minimal" *context of sense and sensibility*. Obviously, the idea is also to explicitly relate the cognitive structure of human senses (visualized in biological terms) to the elementary inventory of representational techniques as actually applied by humans in everyday (and scientific) practise. Also, one would expect that the techniques actually utilized in that practise point to detailed cognitive aspects of selecting an adequate world-view, in the first place.

## 3. Cognitive Foundations of Synergetics

This problem is being analyzed recently with a view to synergetics. The same problem can be shown to be related to the interdisciplinary founding of synergetics itself: namely in terms of the observer and the process of observation and representation. One of the best cognitive formulations of this problem has been given by M. K. Mamardashvilly. [27] "Combination of various problems and sciences, - Mamardashvilly writes, - resulted in the appearance of the utmost necessity to find a more or less precise definition for the notion of *observation*". At least as much precise that it would approach, in its accuracy and explanation, to the accuracy of mathematical and physical notions. In other words, having become the main notion in the theoretical-cognitive structure of physics, as well as, naturally, in the structure of linguistics and other fields, the notion of "observation" does not only depend on the results of investigations of a consciously perceived series of phenomena, but it also requires from psychology or any other science a complete theory of consciousness. This aspect has been discussed in detail at other places. [28] Hence, the metaphysical context of Prigogine's research program, for instance, is to restore the coherence (in some topological sense) of temporal experience represented in its fundamental divisions and oppositions of external and internal, subjective and objective, in order to comprehend the notion of "arrow of time" as the pattern of distinction of events. [29]

This is where the interdisciplinarity of the research problem actually starts. The methodology of its grasping is of a horizontal kind, as E. Laslo says, pointing to a transdisciplinary reconstruction of reality. This may be visualized as a holistic *structurization* of reality, where creativity, polymorphism of languages, metaphor and analogy, network thinking, circular causality and so forth are characteristically in the focus. Here the most important thing is the course taken from the knowledge as " to know how ", rather than from the knowledge as " to know that ". So at this stage of modeling it is mathematics appearing – the language of interdisciplinary dialogue with a rational nucleus.

In fact, the discourse is already one about *compound binary events*: in the theory of relativity it is the measurement of intervals, and in quantum mechanics it is the interconditionality of simultaneous measurements. Binary events or paired acts of measurement are actually relative to the means of supervision. Physical reality is thus allocated with an elementary communication procedure which refers to the concept of connectivity, which is contextual in the sense that it depends on the means of supervision themselves, and hence, it actually dislocates the atomic event in a non-trivial manner.

In a broad sense, *event* assumes: something has taken place, was held, has come true, began to exist and up to that time did not exist. And at the same time the event happens to be elementary, atomic; and it happens to be significant, powerful, epoch-making. Any event may thus be understood in terms of the qualities listed above, depending on a given context. In a sense, with a view to the concept of dislocation, the event tears the fabric of time here and now, but time itself reconciles the event with the life of the past and future by infinitely many strings visualized as contexts. Basically, synergetics assumes plurality and ambiguity of the actual "reopening" of space and time. Hence, it puts forward the main features of the synergetical discourse as a discourse of sciences. So also *senses* arise, as a contextual dislocation of atomic events. In a manner of speaking, it is possible to say that senses refer to the event in a polycontextual manner. Finally, *context* starts from circumstances of place and action, but then expands by loops of conditional sentences, isolating from all conceivable circumstances the new details. This leaves an opening for semantic pluralism, which sprouts, on boundaries and edges of stipulated spaces and ways. And such ambiguity is inevitably connected with an information finiteness of the person which is realized on epistemological borders in any experimental science, but in our case it is due to the technology of judgement, finiteness of depth of any context - one of the aspects of a principle of observability, the attempt of supervision over the infinite whole by its final part. However, axiomatic theories build the system as a "tower" upon the final number of axioms, and usually hope on a final (probably algorithmical) depth of a context, but there are also insurmountable complexities which ask for further analysis.

The diagrammatic language in physics has arisen because of the need of the description of very complex systems. Here is one more reason for which many people from the humanities rejected the classical scientific methodology – they argued in favour of a different level of complexity within the objects of their research, which demanded for different methods. Today we see the obvious rapprochement of positions on the ground of modeling in terms of cognitive diagrams. So, one of the authors showed that the language of the modern quantum theory of a field (in terms of Feynman's diagrams) is structurally isomorphic to inducing grammars of all natural languages - Chomsky's grammars. There is a number of works dealing with the problems alluded to here. [30]

## 3.  Conclusions

Hence, the various topics discussed in this present paper may have pointed sufficiently to a number of possible applications in order to work in the vicinity of the boundary between sciences and humanities. Two immediate aspects of this: the origin and state of natural laws on the one hand, and the mediation of nature and aesthetics on the other, will be topic of a subsequent paper.

## 4.  Acknowledgements

The authors would like to thank the INTAS/NIS cooperation organization of the European Commission at Brussels for financial support under contract number MP/CA 2000-298. We thank also the other members in the various subtasks of this cooperation, in particular Wolfgang Hofkirchner of the University of Technology, Vienna, as chief organizer.